\documentclass{IEEEtran}
\usepackage{cite}
\usepackage{amsmath,amssymb,amsfonts}
\usepackage{graphicx}
\usepackage{textcomp}
\usepackage{subfigure}
\usepackage{caption}

\usepackage{algorithmic}

\usepackage{mathrsfs}
\usepackage{amsmath}
\usepackage{bm}
\usepackage{stfloats}

\def\BibTeX{{\rm B\kern-.05em{\sc i\kern-.025em b}\kern-.08em
    T\kern-.1667em\lower.7ex\hbox{E}\kern-.125emX}}
\begin{document}

\title{The New Purity and Capacity Models for the OAM-mmWave Communication Systems under Atmospheric Turbulence}

\author{Hanqiong Lou, Xiaohu Ge, \IEEEmembership{Senior Memeber, IEEE}, Qiang Li, \IEEEmembership{Memeber, IEEE}}

\maketitle

\begin{abstract}
The orbital angular momentum (OAM) wireless communication technology is widely studied in recent literatures. But the atmospheric turbulence is rarely considered in analyzing the capacity of OAM-based millimeter wave (OAM-mmWave)  communication systems. The OAM-mmWave propagated in the atmosphere environments is usually interfered by the atmospheric turbulence, resulting in the crosstalk among OAM channels, capacity degradation, etc. By taking into account the atmospheric turbulence effect, this paper proposes a new purity model and a new capacity model for the OAM-mmWave communication systems. Simulation results indicate that the OAM-mmWave propagation in the atmosphere environments is evidently interfered by atmospheric turbulence, where the capacity of the OAM-mmWave communication systems decreases with the increase of the transmission frequency.
\end{abstract}


\section{Introduction}
\label{sec:introduction}
\IEEEPARstart{T}{he} electromagnetic waves have the linear momentum in the propagation directions and the angular momentum in the vertical planes of propagation directions. The angular momentum is classified by two types of angular moments, \emph{i.e.}, the spin angular momentum (SAM) describing the electromagnetic wave polarization and the orbital angular momentum (OAM) describing the electromagnetic wave spiral phase\cite{Yao}. The SAM has only two mutually orthogonal states, the OAM has an infinite number of orthogonal states that can be multiplexed along the same propagation axis. Thus, the OAM can be used for improving the signal transmission efficiency without additional frequency band occupancy \cite{Niemiec}.

When the OAM technology is adopted for optical communications, the transmission capacity has been obviously improved \cite{Rusch,Zhu,Tudor,Kai}. However, in the presence of atmospheric turbulence, the phase of optical waves carrying OAM could be changed, which decreases the capacity of OAM wireless communication systems seriously \cite{Tian,Chen,Li}. The propagation characteristics of OAM light waves considering atmospheric turbulence have been investigated in \cite{Sun,Anguita}. Due to the atmospheric turbulence, the energy of the transmitted OAM state spread into neighboring OAM states with different probabilities, which resulted in crosstalk in OAM-Multiplexed free space optical communication systems \cite{Sun}. The crosstalk model of OAM-Multiplexed free space optical communication systems with the atmospheric turbulence effect was used for selecting the optimal set of OAM states to maximize the capacity of OAM communication systems \cite{Anguita}. The spiral spectrum of OAM light waves in Kolmogorov turbulence was researched in \cite{Zhou}. Then the spiral spectrum of OAM beam in non-Kolmogorov turbulence was also studied in \cite{Jiang}.

In recent years, the OAM technology has been used to improve the capacity of wireless communications \cite{Cang,Shin,Kim,Ren}. However, the impact of the atmosphere effect on the OAM wireless communications using radio is rarely investigated\cite{Yuan11,Zhang,Ge,Gao,Wang}. Therefore, it is of significant importance to evaluate the impact of atmospheric turbulence on the capacity of OAM wireless communication systems. Considering the directional characteristic of OAM waves, the capacity of OAM-multiplexed communication systems with unaligned transceiver antenna arrays was shown to be smaller than the capacity of OAM-multiplexed communication systems with aligned transceiver antenna arrays \cite{Yuan11}. Because the OAM waves could decrease the spatial correlation of wireless channels, the capacity of OAM-based Multiple-Input Multiple-Output (MIMO) systems was shown to be larger than the capacity of the traditional MIMO systems in free space with weak multipath effect \cite{Zhang}. Moreover, the energy efficiency of OAM spatial modulation millimeter wave communication systems was shown to be higher than the energy efficiency of OAM-based MIMO systems \cite{Ge}, which made OAM spatial modulation millimeter wave communication systems more suitable for transmissions in long distances efficiently. In view of the fact that the divergence of OAM-based radio vortex waves is more significant than the divergence of radio plane waves, a new type of bifocal lens antenna was used for transmitting OAM-based signals to reduce the propagation attenuation \cite{Gao}. However, in these studies, only Gaussian white noise was considered as the external interference factor in the OAM wireless propagation. An interesting problem arises as to whether the influence of atmospheric turbulence on the OAM wave transmissions is the same as the influence of atmospheric turbulence on the OAM millimeter wave (OAM-mmWave) transmissions.

In this paper the impact of atmospheric turbulence on the OAM-mmWave communication systems has been investigated. A new capacity model has been proposed for the OAM-mmWave communication systems considering the atmospheric turbulence effect. The contributions of this paper are summarized as follows:
\begin{enumerate}
  \item A new purity model is proposed for OAM-mmWave communication systems considering the propagation characteristics of OAM-mmWave with the atmospheric turbulence. In general, the atmospheric turbulence causes the power of the transmitted OAM state to spread into the adjacent OAM states, \emph{i.e.}, the crosstalk effect of atmospheric turbulence.
  \item A capacity model is proposed for the OAM-mmWave communication systems based on the atmospheric turbulence. The crosstalk among channels with different OAM states is characterized based on the proposed capacity model.
  \item Conventionally, without considering the atmospheric turbulence, the capacity of OAM-mmWave communication systems is found to be irrelevant to the transmission frequency. By contrast, in this paper simulation results show that the capacity of OAM-mmWave communication systems, in the presence of the atmospheric turbulence, decreases with the increase of the transmission frequency. Moreover, simulation results indicate that the capacity of OAM-mmWave communication systems decreases with the increase of the refractive index structure constant that is proportional to the atmospheric turbulence intensity.
\end{enumerate}

The rest of this paper is organized as follows. In Section II, a new purity model is established for the OAM millimeter waves considering the atmospheric turbulence. In Section III, the capacity model for OAM-mmWave communication systems with the atmospheric turbulence is proposed. In Section IV, the simulation results are analyzed and discussed. Finally, conclusions are drawn in Section V.

\begin{table}
\caption{A list of symbols appear in this paper.}
\label{table}
\setlength{\tabcolsep}{3pt}
\begin{tabular}{|p{65pt}|p{165pt}|}
\hline
Symbol&
Quantity\\
\hline
${u_{p,l}}(r,\phi ,z) $&
field distribution of the OAM wave \\
$(r,\phi ,z)$&
cylindrical coordinate system\\
$z$&
propagation distance\\
$\alpha$&
arbitrary complex constant\\
$i$&
imaginary unit\\
$p$&
radial index\\
$w(z)$&
beam waist radius of the OAM wave\\
${w_0}$&
beam waist radius at z = 0\\
$R(z)$&
phase front radius of curvature of LG beam at receiver\\
${z_R}$&
Rayleigh distance\\
$\lambda$&
wavelength\\
$f$&
frequency\\
$k$&
wave number\\
${l_0},m$&
transmitted OAM state\\
$l,n$&
received OAM state\\
${r_0}$&
spatial coherence radius of the LG beam wave\\
$\Theta$&
curvature parameter of the LG beam at receiver\\
$\Lambda$&
Fresnel ratio of the LG beam at receiver\\
$C_n^2$&
refractive index structure constant\\
$u(r,\phi ,z)$&
field distribution in atmosphere environments\\
${P_{{l_0}}}({l_0},z)$&
purity of the millimeter wave with OAM state ${l_0}$\\
${P_l}({l_0},z),{P_n}(m,z)$&
power weight of the spiral harmonic component\\
$L$&
the number of OAM channels\\
$S$&
transmitted OAM state set\\
${\mathbf{P}}$&
crosstalk matrix of OAM channels\\
$\Upsilon$&
signal to-interference-and-noise ratio matrix\\
${N_0}$&
additive white Gaussian noise power\\
${P_{TX}}$&
transmitted power\\
${\mathbf{p}}$&
bit error rate matrix of OAM channels\\
${{\mathbf{C}}_{\mathbf{L}}}$&
capacity matrix of OAM channels\\
$C$&
capacity of the communication system\\
${C_{ideal}}$&
capacity of the ideal communication system\\
\hline
\end{tabular}
\label{tab1}
\end{table}

\section{Purity Model based on Atmospheric Turbulence}
\subsection{System model}

\begin{figure}[!t]
\centering
\includegraphics[width=3in]{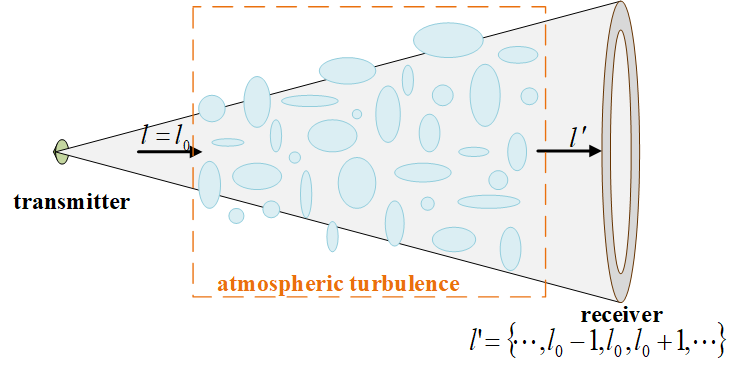}
\caption{An illustration of the OAM-mmWave propagation in the atmosphere environment.}
\label{Fig1}
\end{figure}

As shown in Fig.1, due to the diffractive effect, the propagation of OAM waves is divergent in the free space and the intensity of OAM waves is mainly distributed in the annular region \cite{Anguita}. In this paper the Laguerre-Gaussian (LG) beam is used to describe the OAM wave. Hence, the field distribution of the OAM wave is expressed in cylindrical coordinate system $\left( {r,\phi ,z} \right)$ as \cite{Xu}
\begin{equation}
\begin{split}
  {u_{p,l}}\left( {r,\phi ,z} \right) =& \alpha \sqrt {\frac{{p!}}{{\pi \left( {p + \left| l \right|} \right)!}}} \frac{1}{{w\left( z \right)}}{\left( {\frac{{\sqrt 2 r}}{{w\left( z \right)}}} \right)^{\left| l \right|}}{e^{ - {{\left( {\frac{r}{{w(z)}}} \right)}^2}}} \\
  &\times L_p^{\left| l \right|}\left( {\frac{{2{r^2}}}{{{w^2}\left( z \right)}}} \right){e^{\frac{{ - i\pi {r^2}}}{{\lambda R\left( z \right)}}}}{e^{i\left( {\left| l \right| + 2p + 1} \right)\zeta \left( z \right)}}{e^{ - il\phi }}
\end{split},\tag{1a}
\end {equation}
with
\begin{align*}
  w(z) & ={{w}_{0}}\sqrt{1+{{\left( \frac{z}{{{z}_{R}}} \right)}^{2}}},\tag{1b} \\
  R(z) & =z\left[ 1+{{\left( \frac{\pi w_{0}^{2}}{\lambda z} \right)}^{2}} \right], \tag{1c}
\end{align*}
where $r$ is radial distance, $\phi$ is azimuthal angle, $z$ is propagation distance, $\alpha$ is an arbitrary complex constant, $i$ is an imaginary unit, $p$ is the radial index, $p=0$ is generally configured for OAM systems. $l$ is the OAM state, $l\in \left\{ +|l|,-|l| \right\}$. The rotation direction of the spiral wavefront is determined by the symbol of $l$. The absolute value of $l$ represents the number of equiphase surfaces wound on the propagation axis. When $l=0$, the electromagnetic wave is a conventional plane wave or spherical wave. $w(z)$ is the beam waist radius of the OAM electromagnetic wave at the propagation distance $z$. ${{w}_{0}}$ is the beam waist radius at $z=0$. $\zeta (z) = \arctan \left( {\frac{z}{{{z_R}}}} \right)$ is the Gouy phase, where ${{z}_{R}}=\frac{\pi w_{0}^{2}}{\lambda }$ is the Rayleigh distance, $\lambda$ is the wavelength. $f$ is the frequency. $L_{p}^{|l|}\left( \frac{2{{r}^{2}}}{{{w}^{2}}(z)} \right)$ is a generalized Laguerre polynomial, $L_{p}^{|l|}\left( \frac{2{{r}^{2}}}{{{w}^{2}}(z)} \right)=1$ when $p=0$. The propagation characteristics of OAM waves in the millimeter wave band are studied in this paper. The equiphase surface of the LG beam is a spiral with the curvature radius $R(z)$, which is formulated by (1c). The intensity of OAM waves is mainly distributed in an annular region, where the radius of the ring with the maximum intensity is derived by \cite{Wang}
\[{r_{\max }}(z) = \sqrt {\frac{{\left| l \right|}}{2}} w(z).\tag{2} \]
This is a key factor that directly affects the choice of receiver radius. The radius of receiver is large than ${r_{\max }}(z)$. Because the wavelength of the millimeter wave is longer than the wavelength of light, based on (2), the divergence degree of the OAM-mmWave is larger than the divergence degree of the OAM light wave at the same distance. Due to the practical consideration of the receiver, the propagation distance of the OAM-mmWave is limited.

In the actual atmosphere environments, the atmospheric turbulence interferes with the propagation of OAM-mmWaves and causes the power of the emitted OAM states to spread into the adjacent OAM states. After the OAM-mmWave with the OAM state ${l_0}$ is transmitted in atmosphere environments, the OAM-mmWaves with the OAM state $l \in \left\{ { \cdots ,{l_0} - 1,{l_0},{l_0} + 1, \cdots } \right\}$ are received at the receiver. In this paper, the Kolmogorov turbulence model is used to approximate the atmospheric turbulence interferences in OAM-mmWave propagation. Considering the Kolmogorov turbulent flow, the spatial coherence radius of the LG beam wave is expressed as \cite{LC}
\[{r_0} = {\left[ {\frac{{\rm{8}}}{{{\rm{3}}\left( {a + 0.618{\Lambda ^{\frac{{11}}{6}}}} \right)}}} \right]^{\frac{3}{5}}}{(1.46C_n^2{k^2}z)^{ - \frac{3}{5}}},\tag{3a}\]
with
\begin{align*}
  a & = \frac{{1 - {\Theta ^{\frac{8}{3}}}}}{{1 - \Theta }},\tag{3b} \\
  \Lambda & = \frac{{2z}}{{k{\omega ^2}(z)}},\tag{3c}
\end{align*}
where $k = 2\pi /\lambda $ is the wave number, $\Theta  = 1 + \frac{z}{{R(z)}}$ is the curvature parameter of the LG beam at receiver, $\Lambda$ is the Fresnel ratio of the LG beam at receiver, $C_n^2$ is the refractive index structure constant. Turbulence strength increases with the increase of $C_n^2$ \cite{Anguita}. In near-surface environments, $C_n^2$ is affected by temperature and humidity. The key influencing factor of $C_n^2$ in the optical band is temperature. The key influencing factor of $C_n^2$ in the millimeter wave band is humidity \cite{Mcmillan}, \cite{Hill}. Hence, the value of $C_n^2$ is related to the frequency when the environmental conditions are the same. The $C_n^2$ of the millimeter waves is two orders of magnitude larger than the $C_n^2$ of the light waves \cite{Mcmillan}. When the millimeter wave band is adopted for OAM propagations, the statistical average of $C_n^2$ is $4.0 \times {10^{ - 12}}{{\text{m}}^{{\text{ - }}\frac{{\text{2}}}{{\text{3}}}}}$ in near-surface environments \cite{Hill}.

At present, the research on OAM-mmWave communication systems generally only considers white Gaussian noise. In this case, the orthogonality of OAM-mmWave states is maintained for the propagation of OAM-mmWave \cite{Wang} \cite{Xu}. But the atmospheric turbulence causes the phase distortion and the loss of orthogonality in the actual atmosphere environments. In order to describe this phenomenon theoretically, it is necessary to calculate the purity model of OAM millimeter wave according to the Kolmogorov turbulence model.

\begin{figure*}[bp]
\normalsize
\hrulefill
\begin{align*}
{D_l}({l_0},z) &= \int_0^R  {\left\langle {{{\left| {{a_l}(r,z)} \right|}^2}} \right\rangle rdr}  \hfill \\
&= \frac{1}{{2\pi }}\int_0^R  {\int_0^{2\pi } {\int_0^{2\pi } {\left\langle {u(r,\phi ,z){u^*}(r,\phi ',z)} \right\rangle } {e^{il(\phi  - \phi ')}}d\phi d\phi '} rdr}  \hfill \\
&= \frac{1}{{2\pi }}\int_0^R  {\int_0^{2\pi } {\int_0^{2\pi } {{u_{p,{l_0}}}(r,\phi ,z)u_{p,{l_0}}^*(r,\phi ',z)\left\langle {{e^{\psi (r,\phi ,z) + {\psi ^*}(r,\phi ',z)}}} \right\rangle {e^{il(\phi  - \phi ')}}d\phi } d\phi '} rdr}  \hfill \\
&= \frac{{{\alpha ^2}{w^{ - 2}}(z)p!}}{{2{\pi ^2}(p + \left| {{l_0}} \right|)!}}\int_0^R  {{{\left( {\frac{{2{r^2}}}{{{w^2}(z)}}} \right)}^{\left| {{l_0}} \right|}}{e^{ - \frac{{2{r^2}}}{{{w^2}(z)}} - \frac{{2{r^2}}}{{r_0^2}}}}{{\left[ {L_p^{\left| {{l_0}} \right|}\left( {\frac{{2{r^2}}}{{{w^2}(z)}}} \right)} \right]}^2}\int_0^{2\pi } {\int_0^{2\pi } {{e^{i(l - {l_0})(\phi  - \phi ')}}{e^{\frac{{2{r^2}\cos (\phi  - \phi ')}}{{r_0^2}}}}d\phi } d\phi '} rdr}  \hfill \\
&= \frac{{2{\alpha ^2}{w^{ - 2}}(z)p!}}{{(p + \left| {{l_0}} \right|)!}}\int_0^R  {{{\left( {\frac{{2{r^2}}}{{{w^2}(z)}}} \right)}^{\left| {{l_0}} \right|}}{e^{ - \frac{{2{r^2}}}{{{w^2}(z)}} - \frac{{2{r^2}}}{{r_0^2}}}}{{\left[ {L_p^{\left| {{l_0}} \right|}\left( {\frac{{2{r^2}}}{{{w^2}(z)}}} \right)} \right]}^2}{I_{l - {l_0}}}\left( {\frac{{2{r^2}}}{{r_0^2}}} \right)rdr}  \hfill \tag{8b}   \\
{P_l}({l_0},z) &= \frac{{{D_l}({l_0},z)}}{{{D_{initial}}}} = \frac{{\int_0^R  {{{\left( {\frac{{2{r^2}}}{{{w^2}(z)}}} \right)}^{\left| {{l_0}} \right|}}{e^{ - \frac{{2{r^2}}}{{{w^2}(z)}} - \frac{{2{r^2}}}{{r_0^2}}}}{{\left[ {L_p^{\left| {{l_0}} \right|}\left( {\frac{{2{r^2}}}{{{w^2}(z)}}} \right)} \right]}^2}{I_{l - {l_0}}}\left( {\frac{{2{r^2}}}{{r_0^2}}} \right)rdr} }}{{\frac{{{w^2}(z)}}{{w_0^2}}\int_0^R  {{{\left( {\frac{{2{r^2}}}{{w_0^2}}} \right)}^{\left| {{l_0}} \right|}}{e^{ - \frac{{2{r^2}}}{{w_0^2}}}}{{\left[ {L_p^{\left| {{l_0}} \right|}\left( {\frac{{2{r^2}}}{{w_0^2}}} \right)} \right]}^2}rdr} }}\tag{11}  \\
{P_{{l_{\text{0}}}}}({l_0},z) &= \frac{{{D_{{l_{\text{0}}}}}({l_0},z)}}{{{D_{initial}}}} = \frac{{\int_0^R  {{{\left( {\frac{{2{r^2}}}{{{w^2}(z)}}} \right)}^{\left| {{l_0}} \right|}}{e^{ - \frac{{2{r^2}}}{{{w^2}(z)}} - \frac{{2{r^2}}}{{r_0^2}}}}{{\left[ {L_p^{\left| {{l_0}} \right|}\left( {\frac{{2{r^2}}}{{{w^2}(z)}}} \right)} \right]}^2}{I_{\text{0}}}\left( {\frac{{2{r^2}}}{{r_0^2}}} \right)rdr} }}{{\frac{{{w^2}(z)}}{{w_0^2}}\int_0^R  {{{\left( {\frac{{2{r^2}}}{{w_0^2}}} \right)}^{\left| {{l_0}} \right|}}{e^{ - \frac{{2{r^2}}}{{w_0^2}}}}{{\left[ {L_p^{\left| {{l_0}} \right|}\left( {\frac{{2{r^2}}}{{w_0^2}}} \right)} \right]}^2}rdr} }}\tag{12}
\end{align*}
\end{figure*}

\subsection{Purity model}

In the actual atmosphere environments, the atmospheric turbulence causes the phase distortion and intensity fading in OAM-mmWave propagation \cite{Amphawan}. Based on the measured results, the value of intensity fading is small in most atmospheric turbulence scenarios \cite{YZhang}, \cite{Paterson}. Considering that this paper focuses on the phase distortion due to the atmospheric turbulence, the intensity fading is ignored in this paper. In this case, the orthogonality of OAM-mmWave states can't be maintained for the propagation of OAM-mmWave in atmosphere environments. Therefore, it is necessary to analyze the propagation characteristics of OAM-mmWave with the atmospheric turbulence.

Applying the Rytov approximation, the phase distortion of spherical wave caused by the atmospheric turbulence is expressed as \cite{LC}
\[u(r,\phi ,z) = {u_{p,{l_0}}}(r,\phi ,z){e^{\psi (r,\phi ,z)}}.\tag{4a}\]
Based on the quadratic approximation \cite{LC}, $\psi (r,\phi ,z)$ satisfies
\[\left\langle {{e^{\psi (r,\phi ,z) + {\psi ^*}(r,\phi ',z)}}} \right\rangle  = {e^{\frac{{2{r^2}}}{{r_0^2}}\left( {\cos (\phi ' - \phi ) - 1} \right)}}.\tag{4b}\]
Because multiple spherical waves with different phases can be used to generate an OAM wave \cite{YRY}, \cite{SMM}, the Rytov approximation and quadratic approximation of spherical waves can be used for OAM waves.

When the atmospheric turbulence is considered for the OAM wave propagation, the power of the emitted OAM state spreads into the adjacent OAM states \cite{Yuan10}. ${e^{ - il\phi }}$ represents the helical phase distribution, so (1a) can be written as
\[{u_{p,l}}\left( {r,\phi ,z} \right) = \frac{1}{{\sqrt {2\pi } }}{a_l}(r,z){e^{ - il\phi }},\tag{5a}\]
with
\begin{equation}
\begin{split}
    {a_l}\left( {r,z} \right) =& \alpha \sqrt {\frac{ 2 {p!}}{{ \left( {p + \left| l \right|} \right)!}}} \frac{1}{{w\left( z \right)}}{\left( {\frac{{\sqrt 2 r}}{{w\left( z \right)}}} \right)^{\left| l \right|}}{e^{ - {{\left( {\frac{r}{{w(z)}}} \right)}^2}}} \\
    &\times L_p^{\left| l \right|}\left( {\frac{{2{r^2}}}{{{w^2}\left( z \right)}}} \right){e^{\frac{{ - i\pi {r^2}}}{{\lambda R\left( z \right)}}}}{e^{i\left( {\left| l \right| + 2p + 1} \right)\zeta \left( z \right)}}
\end{split}.\tag{5b}
\end {equation}
In this case the field strength of the OAM-mmWave in (4a) is expanded into multiple OAM states (The OAM states interval is $\left( { - \infty , + \infty } \right)$). Therefore, (4a) can be rewritten as
\[u(r,\phi ,z) = \frac{1}{{\sqrt {2\pi } }}\sum\limits_{l =  - \infty }^\infty  {{a_l}(r,z){e^{ - il\phi }}}.\tag{6} \]
According to the discrete-time Fourier transform, the coefficient ${a_l}(r,z)$ is expressed as
\[{a_l}(r,z) = \frac{1}{{\sqrt {2\pi } }}\int_0^{2\pi } {u(r,\phi ,z){e^{il\phi }}d\phi }.\tag{7} \]
The power weight of the spiral harmonic component with OAM state $l$ is expressed as
\[{P_l}({l_0},z) = {D_l}({l_0},z)/\sum\limits_{n =  - \infty }^\infty  {{D_n}({l_0},z)},\tag{8a} \]
with ${D_l}({l_0},z)$ given in (8b), where $\left\langle {{{\left| {{a_l}(r,z)} \right|}^2}} \right\rangle $ represents the probability density distribution of OAM-mmWave with state $l$, $\int_0^{2\pi } {{e^{ - in\phi ' + \eta \cos (\phi ' - \phi )}}d\phi ' = 2\pi {e^{ - in\phi }}{I_n}(\eta )}$ \cite{IS}, ${I_n}( \cdot )$ is the modified n-order Bessel function of the first kind, $R$ is the radius of receiver. Purity is defined as the proportion of the original state power in the total power of the OAM-mmWaves after the propagation in atmosphere environments \cite{Yin}. From the definition of purity, it is known that ${P_{{l_0}}}({l_0},z)$ is the purity of the millimeter wave with OAM state ${l_0}$.

The undistorted electric field of (1a) at $z = 0$ is given as
\[{u_{p,{l_0}}}(r,\phi ,0) = \frac{1}{{\sqrt {2\pi } }}{\beta _{{l_0}}}(r,0){e^{ - i{l_0}\phi }}.\tag{9} \]
When $z = 0$, $\sum\limits_{n =  - \infty }^\infty  {{D_n}({l_0},z)} $ is denoted as the ${D_{initial}}$, which is expressed by
\begin{equation}
\begin{split}
  {D_{initial}} =& \int_0^R  {\left\langle {{{\left| {{\beta _{{l_0}}}(r,0)} \right|}^2}} \right\rangle rdr}  \hfill \\
   =& 2{\alpha ^2} \times \frac{{p!}}{{(p + \left| {{l_0}} \right|)!}} \times \frac{1}{{w_0^2}} \hfill \\
   &\times \int_0^R  {{{\left( {\frac{{2{r^2}}}{{w_0^2}}} \right)}^{\left| {{l_0}} \right|}}{e^{\frac{{ - 2{r^2}}}{{w_0^2}}}}{{\left[ {L_p^{\left| {{l_0}} \right|}\left( {\frac{{2{r^2}}}{{w_0^2}}} \right)} \right]}^2}rdr}  \hfill
\end{split}.\tag{10}
\end {equation}

When the OAM-mmWave power weight is normalized by ${D_{initial}}$ , the power weight of the spiral harmonic component with OAM state $l$ is given by (11). When $l = {l_0}$ in (11), the purity ${P_{{l_0}}}({l_0},z)$ of the millimeter wave with OAM state ${l_0}$ is derived by (12).

In order to make full use of the orthogonality of OAM-mmWave states, OAM-mmWave communication systems usually have multiple OAM waves with different OAM states for propagation simultaneously. Therefore, it is necessary to analysis the signal power crosstalk among adjacent OAM states according to the purity formula and obtain the capacity model of the OAM-mmWave communication system considering atmospheric turbulence.

\section{Capacity Model based on Atmospheric Turbulence}

The orthogonality among OAM states can be used to simultaneously transmit information by the wave of OAM-mmWave communication systems with the same frequency. As a consequent, electromagnetic waves multiplexing with different OAM states can be regarded as different channels in OAM-mmWave communication systems, \emph{i.e.}, an OAM state is regarded as an OAM channel. The OAM state set transmitted by the transmitter is assumed as $S$. When the OAM-mmWave is propagated in atmosphere environments, the atmospheric turbulence results in the signal power crosstalk among adjacent OAM states for OAM-mmWave communication systems. In this case, the received signal power at the $n - th$  OAM channel is not only from the OAM-mmWave with the state $n$  but also from the OAM-mmWaves with different states. When the OAM-mmWave communication system has $L = 2N + 1$ symmetrically distributed OAM channels \emph{i.e.}, $S = \left\{ { - N, - N + 1, \cdots ,0, \cdots ,N - 1,N} \right\}$, the crosstalk matrix of the OAM-mmWave communication system is expressed as (13),
\begin{small}
\[{\mathbf{P}} = \left[ {\begin{array}{*{20}{c}}
  {{P_{ - N}}( - N,z)}& \cdots &{{P_n}( - N,z)}& \cdots &{{P_N}( - N,z)} \\
   \vdots & \ddots & \vdots & \ddots & \vdots  \\
  {{P_{ - N}}(m,z)}& \cdots &{{P_n}(m,z)}& \cdots &{{P_N}(m,z)} \\
   \vdots & \ddots & \vdots & \ddots & \vdots  \\
  {{P_{ - N}}(N,z)}& \cdots &{{P_n}(N,z)}& \cdots &{{P_N}(N,z)}
\end{array}} \right]\tag{13}\]
\end{small}where ${P_n}(m,z)$, $ - N \leqslant n \leqslant N$, $ - N \leqslant m \leqslant N$ is the power weight of the spiral harmonic component with OAM state $n$, when the OAM-mmWave with the OAM state $m$ is transmitted in atmosphere environments. When $l = n$, ${l_0} = m$, $m \ne n$ are substituted into (11), ${P_n}(m,z)$ represents that the normalized power is spread from the state $m$ into the state $n$. The $m - th$ row of the matrix ${\mathbf{P}}$ describes that the normalized wave power of the $m - th$ OAM channel is spread into OAM channels which are included in the set $S$. The $n - th$  column of the matrix ${\mathbf{P}}$ includes the desired wave power of $n - th$ OAM channel and the spread wave power from other OAM channels which are included in the set $S$. Based on each column of the crosstalk matrix ${\mathbf{P}}$ , the signal-to-interference-and-noise ratio (SINR) of OAM channels is expressed as
\[\Upsilon = [\begin{array}{*{20}{c}}
  {{\gamma _{ - N}}}& \cdots &{{\gamma _n}}& \cdots &{{\gamma _N}}
\end{array}],\tag{14a} \]
with
\[{\gamma _n} = \frac{{{P_n}(n,z)}}{{\sum\limits_{m \in S}^{m \ne n} {{P_n}(m,z)}  + \frac{{{N_0}}}{{{P_{TX}}}}}},\tag{14b} \]
where ${\gamma _n}$ is the SINR of the $n - th$ OAM channel, ${N_0}$ is the additive white Gaussian noise power, ${P_{TX}}$ is the transmitted power. When the OAM-mmWave communication systems apply the orthogonal modulation, the bit error rate of OAM channels is derived as
\[{\mathbf{p}} = \left[ {\begin{array}{*{20}{c}}
  {{p_{ - N}}}& \cdots &{{p_n}}& \cdots &{{p_N}}
\end{array}} \right],\tag{15a} \]
with \cite{Anguita}
\[{p_n}{\text{ = }}\frac{1}{2}erfc\left( {\sqrt {\frac{{{\gamma _n}}}{2}} } \right),\tag{15b} \]
where ${p_n}$ is the bit error rate of the $n - th$ OAM channel, $erfc( \cdot )$ is the complementary error function. It must be noted that the calculation of the bit error rate is related to the modulation formats. Moreover, even the same kind of modulation format such as M-QAM has different bit error rate when the M takes different values. 5-QAM and 9-QAM have better crosstalk tolerance than regular 4-QAM and 8-QAM, because 5-QAM and 9-QAM contains zero-power symbols which do not interfere with the symbols of adjacent channels \cite{Djordjevic}. The capacity of OAM channels is derived by
\[{{\mathbf{C}}_{\mathbf{L}}} = \left[ {\begin{array}{*{20}{c}}
  {C({p_{ - N}})}& \cdots &{C({p_n})}& \cdots &{C({p_N})}
\end{array}} \right],\tag{16a} \]
with \cite{Anguita}
\[C({p_n}) = 1 + {p_n}{\log _2}{p_n} + (1 - {p_n}){\log _2}(1 - {p_n}).\tag{16b}\]
With the assumption that each OAM channel is a binary symmetric channel, $C({p_n})$ is the capacity of the $n - th$ OAM channel. Based to (14b), $\sum\limits_{m \in S}^{m \ne n} {{P_n}(m,z)} $ which represents the normalized wave power spread from other OAM channels is configured as independent each other. Therefore, the capacity of the OAM-mmWave communication system based on atmospheric turbulence is \cite{Tian}
\[C = \sum\limits_{n \in S} {C({p_n})},\tag{17} \]
which is the row sum of the matrix ${{\mathbf{C}}_{\mathbf{L}}}$ in (16a).

In the ideal free space, there is no crosstalk among OAM channels with different OAM states, \emph{i.e.}, ${P_n}(m,z) = 0$ when $n \ne m$ and ${P_n}(m,z) = 1$ when $n = m$. In this case, the SINR of each OAM channel satisfies $\gamma  = {P_{TX}}/{N_0}$. As a consequent, the capacity of the ideal OAM-mmWave communication system without atmospheric turbulence is derived as
\begin{small}
\begin{equation}
\begin{split}
  {C_{ideal}} =& \frac{L}{2}erfc\left( {\sqrt {\frac{{{P_{TX}}}}{{2{N_0}}}} } \right){\log _2}\left( {\frac{1}{2}erfc\left( {\sqrt {\frac{{{P_{TX}}}}{{2{N_0}}}} } \right)} \right) \hfill \\
   &+ L + \left( {L - \frac{L}{2}erfc\left( {\sqrt {\frac{{{P_{TX}}}}{{2{N_0}}}} } \right)} \right) \hfill \\
   &\times {\log _2}\left( {1 - \frac{1}{2}erfc\left( {\sqrt {\frac{{{P_{TX}}}}{{2{N_0}}}} } \right)} \right) \hfill
\end{split} .\tag{18}
\end{equation}
\end{small}

\section{Simulation Results and Discussions}

The capacity model of OAM-mmWave communication systems considering the atmospheric turbulence is analyzed in this section. The OAM-mmWave communication system has $L = 2N + 1$ symmetrically distributed OAM states. The OAM state set transmitted by the transmitter is $S{\rm{ = }}\left\{ {{\rm{ - }}N, - N + 1, \cdots ,0, \cdots ,N - 1,N} \right\}$. $SN{R_0} = 10 \times \lg ({P_{TX}}/{N_0})$ is the signal-to-noise ratio. Default values of simulation parameters are set as follows: the beam waist radius ${w_0} = 0.01{\text{m}}$ , the propagation distance $z = 100{\text{m}}$, the transmission frequency $f = 100{\text{GHz}}$, the refractive index structure constant $C_n^2 = 1 \times {10^{ - 12}}{{\text{m}}^{{\text{ - }}\frac{{\text{2}}}{{\text{3}}}}}$, the signal-to-noise ratio $SN{R_0} = 10{\text{dB}}$, the radius of receiver $R = 50{\rm{m}}$.

\begin{table}
\caption{Default values of parameters.}
\label{table}
\setlength{\tabcolsep}{3pt}
\begin{tabular}{|p{165pt}|p{65pt}|}
\hline
default parameter&
value\\
\hline
beam waist radius ${w_0}$&
0.01m \\
propagation distance $z$&
100m\\
transmission frequency $f$&
100GHz\\
refractive index structure constant $C_n^2$&
$1 \times {10^{{\text{ - }}12}}{\text{ }}{{\text{m}}^{ - \frac{{\text{2}}}{{\text{3}}}}}$\\
signal-to-noise ratio $SN{R_0}$&
10dB\\
receiver radius $R$&
50m\\
\hline
\end{tabular}
\label{tab1}
\end{table}

Fig. 2 manifests that the power of the emitted OAM state spreads into the adjacent OAM states. The emitted OAM state is 10. Because of atmospheric turbulence, the power weight of OAM-mmWave with state 10 is 73.67\%. The power weight of OAM-mmWave with state 9 is 12.01\%. The power weight of OAM-mmWave with state 11 is 12.01\%. The power weight of OAM-mmWave with state 8 is 1.08\%. The power weight of OAM-mmWave with state 12 is 1.08\%. The power weight of the other OAM states is very small. Most of the power remains in the emitted OAM state. The closer the OAM state is to the emitted OAM state, the higher its energy weight is.

\begin{figure}[!t]
\centering
\includegraphics[width=3in]{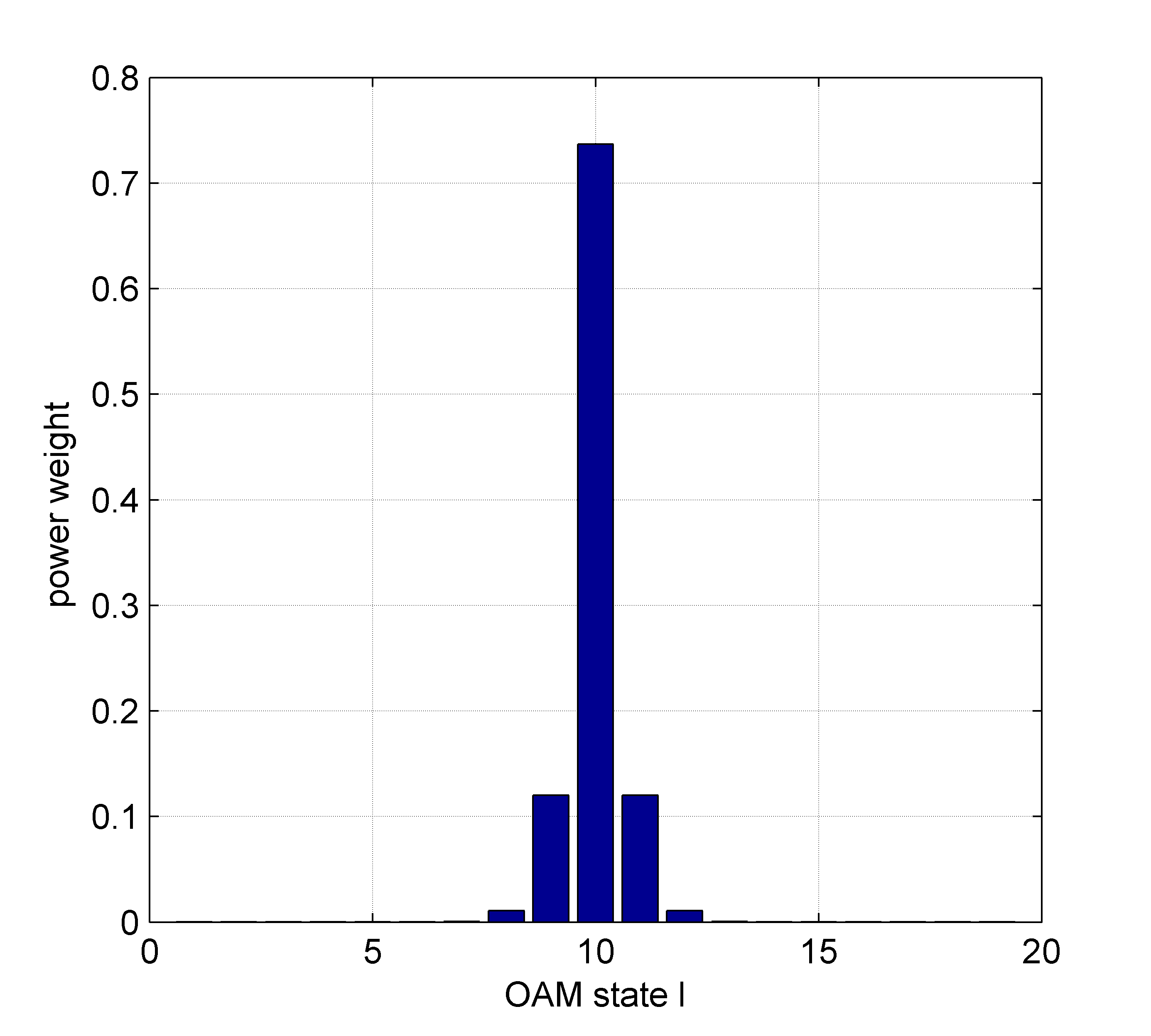}
\caption{The power weight of the emitted OAM state spread into the adjacent OAM states.}
\label{Fig2}
\end{figure}

Fig. 3 illustrates the impact of the number of OAM states on the capacity of OAM-mmWave communication systems with different transmission frequencies. When the transmission frequency effect is ignored in OAM-mmWave communication systems, i.e., the atmospheric turbulence effect is ignored, the capacity of the ideal OAM-mmWave communication system based on (18) is plotted by the black dashed line in Fig. 3. When the transmission frequency is fixed, the capacity of OAM-mmWave communication systems increases with the increase of the number of OAM states. When the number of OAM states is fixed, the capacity of OAM-mmWave communication systems decreases with the increase of the transmission frequency. Moreover, the capacity of OAM-mmWave communication systems considering the atmospheric turbulence is less than the capacity of ideal OAM-mmWave communication systems.

\begin{figure}[!t]
\centering
\includegraphics[width=3in]{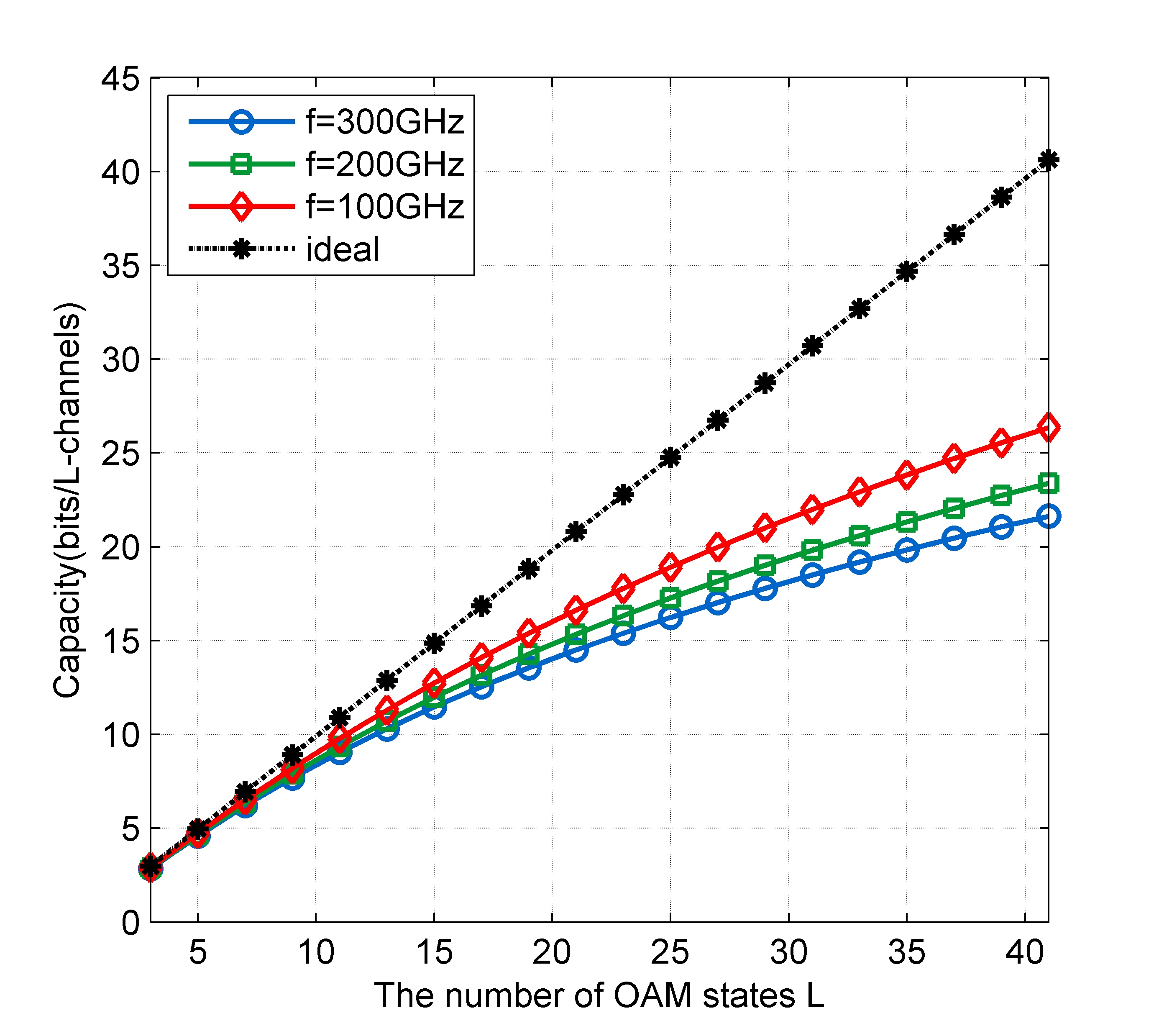}
\caption{Impact of the number of OAM states on the capacity of OAM-mmWave communication systems with different transmission frequencies.}
\label{Fig3}
\end{figure}

Fig. 4 shows the impact of the number of OAM states on the capacity of OAM-mmWave communication systems with different refractive index structure constants. When the number of OAM states is fixed, the capacity of OAM-mmWave communication systems decreases with the increase of the refractive index structure constants which are proportional to the atmospheric turbulence intensity.

\begin{figure}[!t]
\centering
\includegraphics[width=3in]{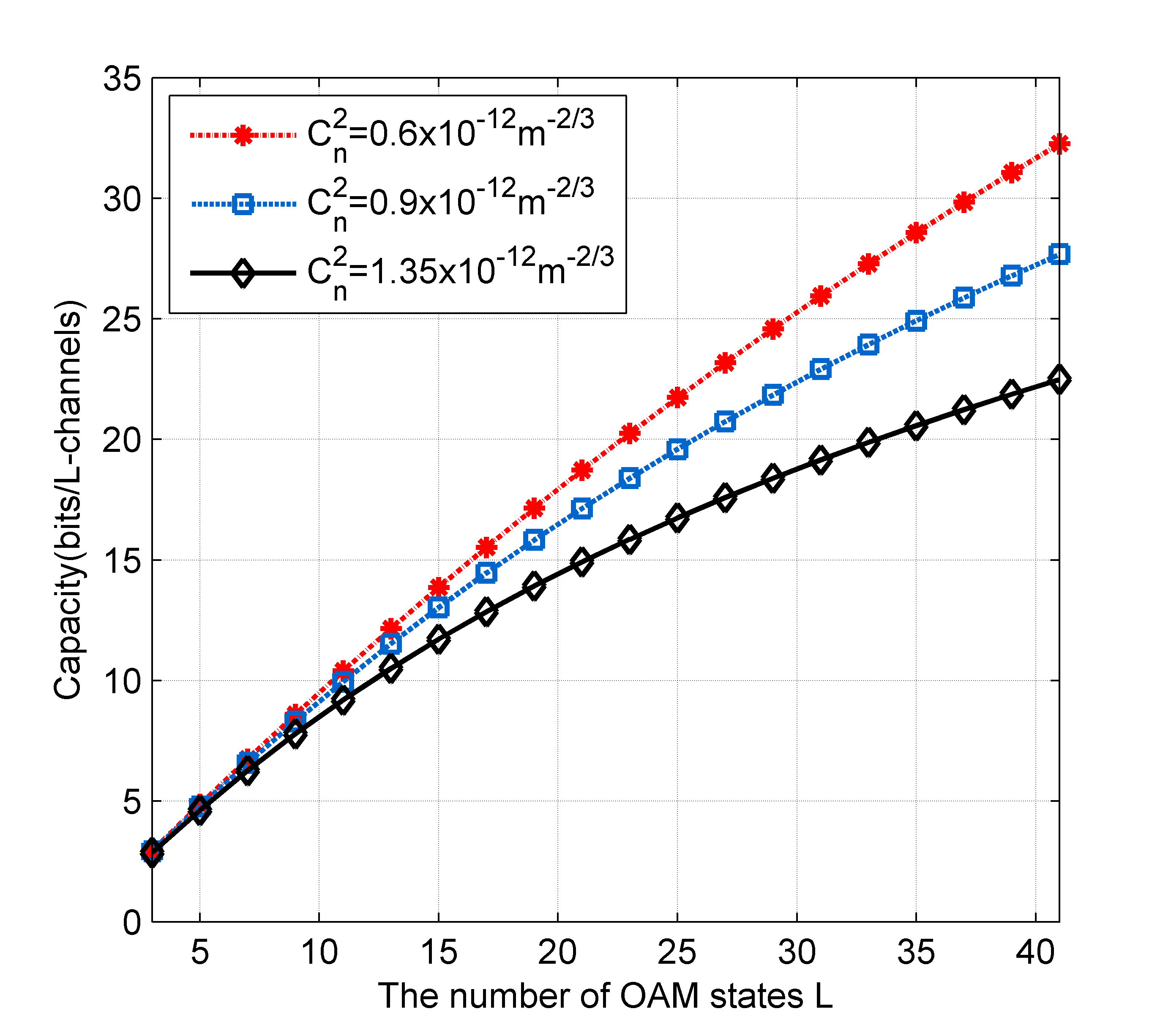}
\caption{Impact of the number of OAM states on the capacity of OAM-mmWave communication systems with different refractive index structure constants.}
\label{Fig4}
\end{figure}

Fig. 5 describes the impact of the number of OAM states on the capacity of OAM-mmWave communication systems with different propagation distances. When the number of OAM states is fixed, the capacity of OAM-mmWave communication systems decreases with the increase of the propagation distance.

\begin{figure}[!t]
\centering
\includegraphics[width=3in]{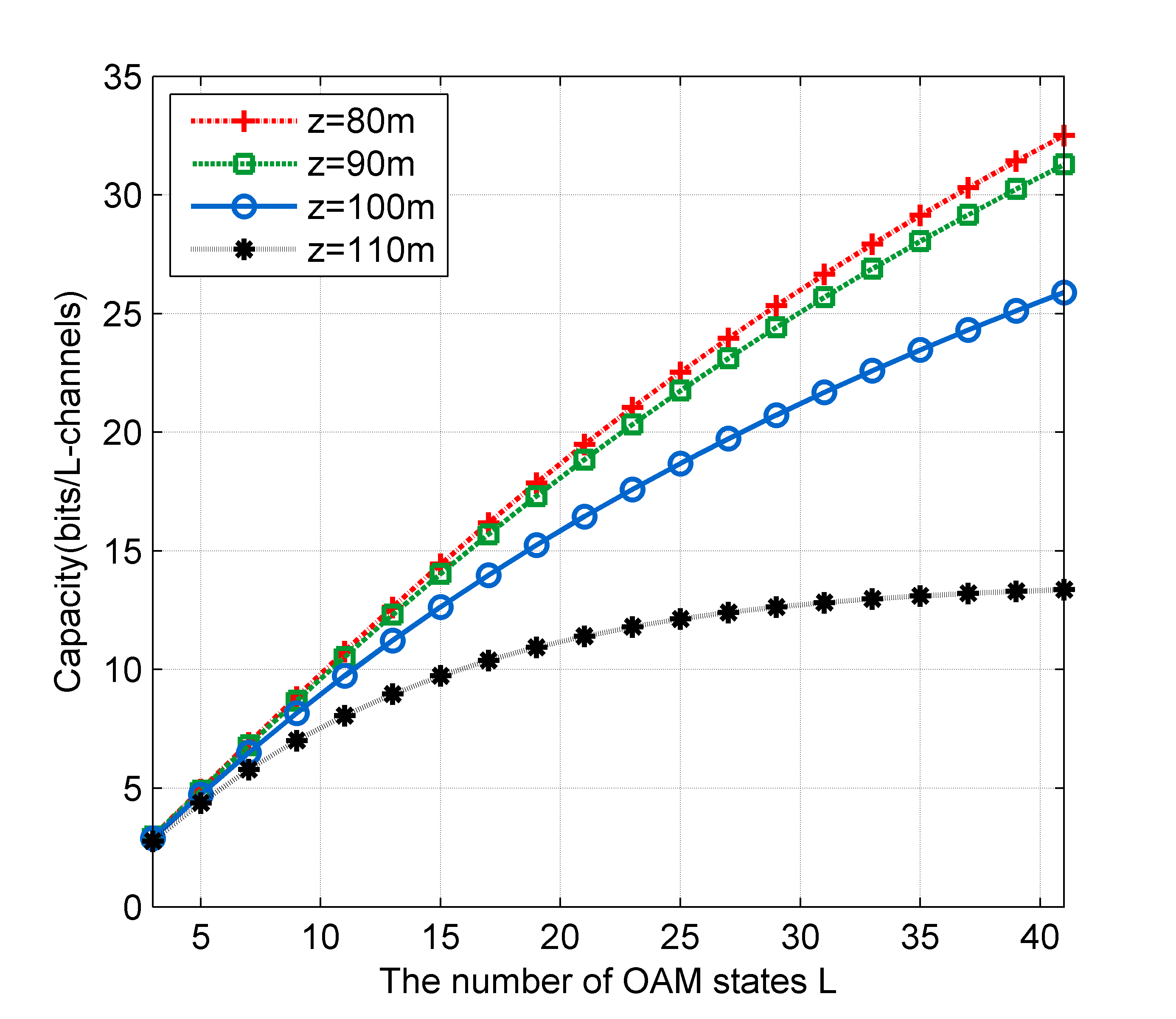}
\caption{Impact of the number of OAM states on the capacity of OAM-mmWave communication systems with different propagation distances.}
\label{Fig5}
\end{figure}

Fig. 6 depicts the impact of the number of OAM states on the capacity of OAM-mmWave communication systems with different $SN{R_0}$ values. When the number of OAM states is fixed, the capacity of OAM-mmWave communication systems increases with the increase of the $SN{R_0}$ values.

\begin{figure}[!t]
\centering
\includegraphics[width=3in]{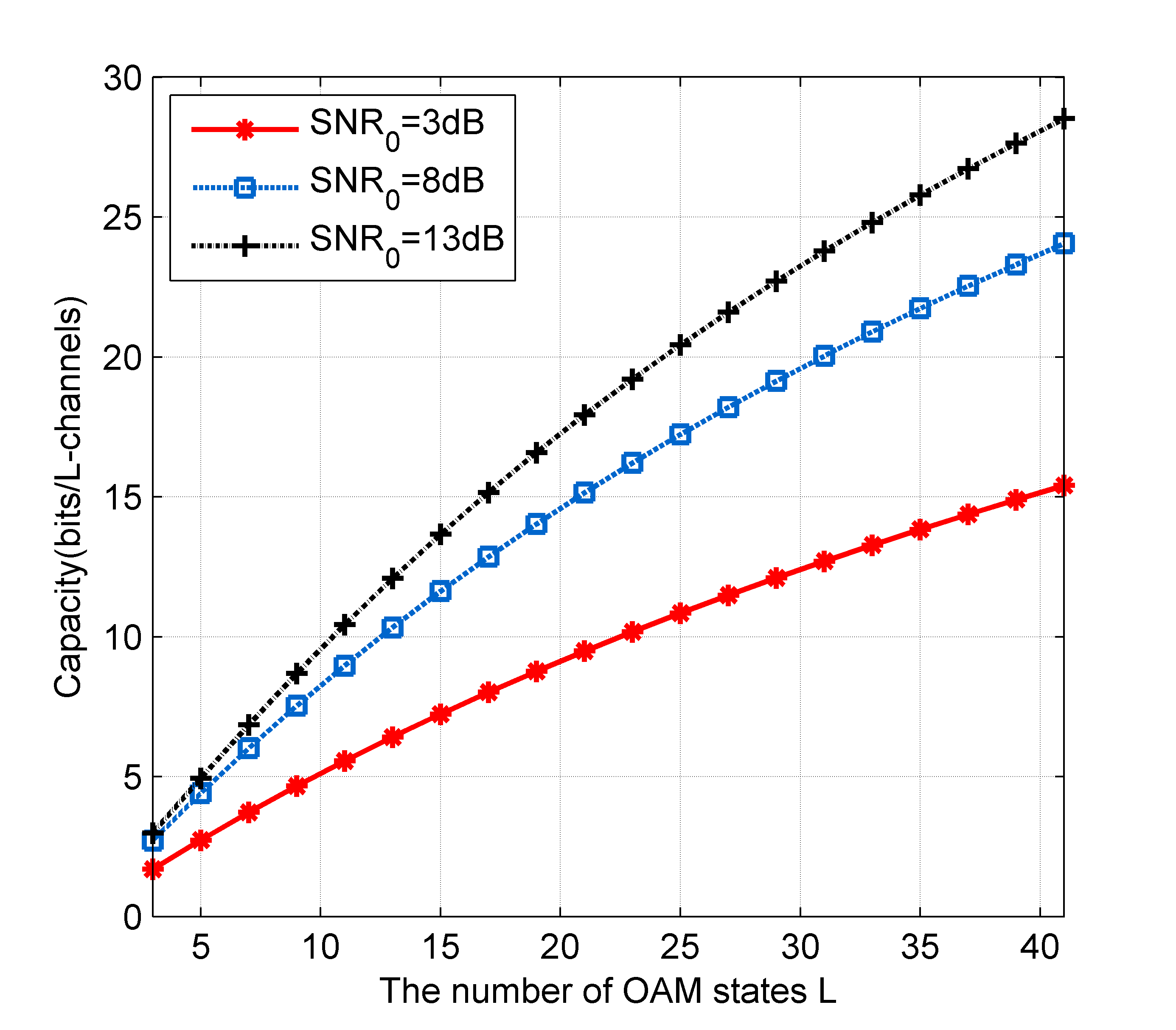}
\caption{Impact of the number of OAM states on the capacity of OAM-mmWave communication systems with different $SN{R_0}$ values.}
\label{Fig6}
\end{figure}

\section{Conclusions}

In conventional studies on OAM wireless communications, the atmospheric turbulence is rarely considered for analyzing the capacity of OAM-mmWave communication systems. By taking into account the effect of atmospheric turbulence, this paper proposes a new purity model and a new capacity model for the OAM-mmWave communication systems. When the atmospheric turbulence effect is ignored in conventional OAM wireless communication systems in most exiting works, the impact of transmission frequency on the OAM-mmWave communication systems is also ignored. However, simulation results of this paper show that the capacity of OAM-mmWave communication systems considering the atmospheric turbulence effect decreases with the increase of the transmission frequency. This paper provides guidelines for designing practical OAM-mmWave communication systems under the influence of atmospheric turbulence. The future research direction of this paper is to analyze the OAM-mmWave communication system model under Non-Kolmogorov turbulence model which is more accurate than Kolmogorov turbulence model to describe the real turbulence environment.

Various methods for mitigating the effect of atmospheric turbulence on OAM waves have been proposed in the optical band, such as MIMO equalization \cite{HHuang}, adaptive optics \cite{YRen}, channel coding \cite{ZQu}. MIMO equalization can effectively improve the OAM signal quality and reduce the bit error rate in atmospheric turbulence. It is more suitable for mitigating the effect of weak atmospheric turbulence. Adaptive optics compensates for the effect of turbulence by using data-carrying Gaussian beacons. But it requires complex optical equipment. Channel coding, such as LDPC, can reduce crosstalk between modes, whether under strong turbulence or weak turbulence. Because no optical equipment is required, MIMO equalization and channel coding are more suitable for the OAM waves in the millimeter band. In order to find the better mitigation method in the millimeter band, comparing MIMO equalization and channel coding will be the future research direction of this paper.

\ifCLASSOPTIONcaptionsoff
  \newpage
\fi

\begin{IEEEbiography}[{\includegraphics[width=1in,height=1.25in,clip,keepaspectratio]{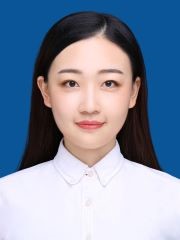}}]{Hanqiong Lou}
received the B.E. degree in communication engineering from Huazhong University of Science and Technology (HUST), Wuhan, China in 2017. She is currently working toward the M.S. degree in HUST. Her research interests mainly in the orbital angular momentum (OAM) technology.
\end{IEEEbiography}

\begin{IEEEbiography}[{\includegraphics[width=1in,height=1.25in,clip,keepaspectratio]{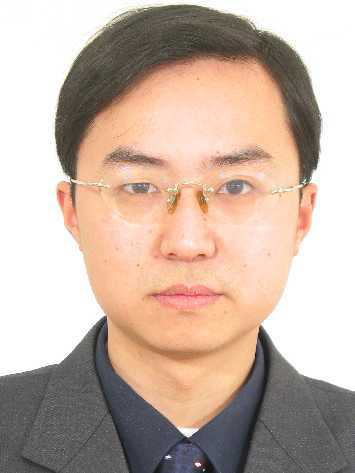}}]{Xiaohu Ge}
(M'09--SM'11) is currently a full Professor with the School of Electronic Information and Communications at Huazhong University of Science and Technology (HUST), China. He is an adjunct professor with the Faculty of Engineering and Information Technology at University of Technology Sydney (UTS), Australia. He received his PhD degree in Communication and Information Engineering from HUST in 2003. He has worked at HUST since Nov. 2005. Prior to that, he worked as a researcher at Ajou University (Korea) and Politecnico Di Torino (Italy) from Jan. 2004 to Oct. 2005. His research interests are in the area of mobile communications, traffic modeling in wireless networks, green communications, and interference modeling in wireless communications. He has published about 200 papers in refereed journals and conference proceedings and has been granted about 25 patents in China. He services as an IEEE Distinguished Lecturer and an Associate Editor for the IEEE ACCESS, IEEE Wireless Communications and IEEE Transactions on Vehicular Technology.
\end{IEEEbiography}

\begin{IEEEbiography}[{\includegraphics[width=1in,height=1.25in,clip,keepaspectratio]{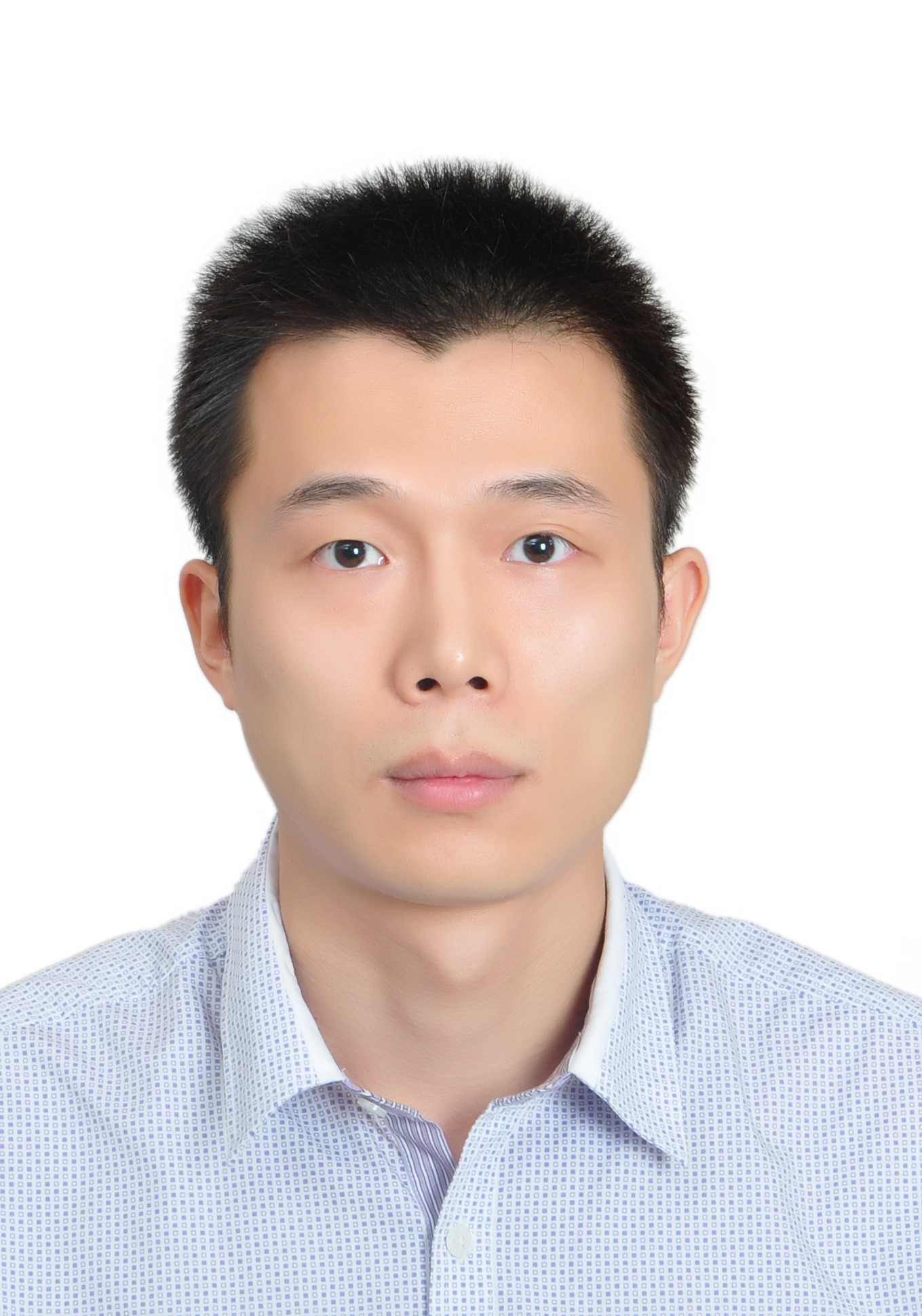}}]{Qiang Li}
(M'16) received the B.Eng. degree in communication engineering from the University of Electronic Science and Technology of China (UESTC), Chengdu, China, in 2007 and the Ph.D. degree in electrical and electronic engineering from Nanyang Technological University (NTU), Singapore, in 2011. From 2011 to 2013, he was a Research Fellow with NTU. From March 2015 to June 2015, he was a visiting scholar with the University of Sheffield, Sheffield, UK. Since 2013, he has been an Associate Professor with Huazhong University of Science and Technology (HUST), Wuhan, China. His current research interests include next generation mobile communications, software-defined networking, cooperative edge caching, cognitive radios/spectrum sharing, simultaneous wireless information and power transfer, wireless cooperative communications and full-duplex relays.
\end{IEEEbiography}

\end{document}